\documentclass[doublespaced,a4paper,12pt]{article}
\usepackage{amssymb}
\usepackage[polish, english]{babel}
\usepackage[cp1250]{inputenc}
\usepackage{amsmath,graphicx,epsf,pstricks,amsfonts,epsfig,latexsym,wasysym}
\title{\bf A relation between the Barbero--Immirzi parameter and the standard model}
\author{Bogus{\l}aw Broda\footnote{bobroda@uni.lodz.pl}\; and\;
Micha{\l} Szanecki\footnote{michalszanecki@wp.pl}\\
\small\textit{Department of Theoretical Physics,}
\small\textit{University of {\L}\'od\'z}\\
\small\textit{Pomorska 149/153, PL--90--236 {\L}\'od\'z, Poland}}
\date{}
\begin{document}
\maketitle
\begin{quote}
\begin{flushleft}
\textbf{Key Words:}\vspace{-.25cm} Ashtekar canonical variables, loop quantum gravity, Barbero--Immirzi \vspace{-.25cm}parameter, Holst action, Nieh--Yan form, ABJ chiral anomaly, Schwinger \vspace{-.25cm}proper-time formalism, Seeley--DeWitt heat-kernel expansion, space-time with torsion.\\
\vspace{0.2cm} \textbf{PACS 2010:}
\vspace{-.25cm}02.40.Hw (Classical differential geometry),
02.40.Ma (Global differential geometry),
\vspace{-.25cm}04.20.Fy (Canonical formalism, Lagrangians, and variational principles),
\vspace{-.25cm}04.60.Pp (Loop quantum gravity, quantum geometry, spin foams),
\vspace{-.25cm}04.70.Dy (Quantum aspects of black holes, evaporation, thermodynamics),
\vspace{-.25cm}11.15.Tk (Other nonperturbative techniques),
11.30.Rd (Chiral symmetries),
\vspace{-.25cm}12.10.Dm (Unified theories and models of strong and electroweak interactions),
12.60.Rc (Composite models). \\
\end{flushleft}
\vspace{0.5cm} It has been shown that Sakharov's induced, from the fields entering the standard model, Barbero--Immirzi parameter $\gamma$ assumes, in the framework of euclidean formalism, the UV cutoff-independent value, 1/9. The calculus uses the Schwinger's proper-time formalism, the Seeley--DeWitt heat-kernel expansion, and it is akin to the derivation of the ABJ chiral anomaly in space-time with torsion.
\end{quote}

The Barbero--Immirzi (BI) parameter $\gamma$ is an a priori free parameter in the framework of the modern approach to canonical gravity (Ashtekar's formalism) \cite{Ashtekar}.  In the Holst extended action for gravity \cite{Holst} the BI parameter $\gamma$ resides in the additional term of the full (Holst) action. One can easily further extend the Holst contribution \cite{Rezende} yielding, in particular, the Nieh--Yan (NY) term (the role that NY invariant plays in gravity has been studied in \cite{Mercuri1}, while an extension to a possible new scenario where BI parameter is promoted to a field, has been studied in \cite{Mercuri2}). Because of topological nature of the NY term, it does not modify classical gravity but it influences quantum theory. (Accidentally, it appears, and we will show it, that, in a sense, also an opposite situation takes place. Namely, the NY term can be quantumly induced by dominant part of one-loop contributions coming from chiral matter fields.)\\
One should stress that there is a known approach using the black-hole entropy to fix the value of the BI parameter $\gamma$ (see, e.g.\ \cite{Domagala}). The main objective of our paper is to show that our method of Sakharov's inducing of the NY term also fix the BI parameter $\gamma$, and moreover it does it in an independent way. We will work in the framework of euclidean formalism applying the Sakharov idea of induced gravity (one-loop dominance) to the standard model of particle physics \cite{Sakharov}. As is well-known, the dominant part of one-loop contributions coming from the fields entering the standard model coupled to gravity (in principle, from any field coupled to gravity) induces (besides the ``cosmological term'') the Einstein--Hilbert (EH) action of (classical) gravity \cite{Sakharov},\cite{Visser},\cite{Broda1}. But there is some exception we are especially interested in. Namely, we will show that chiral fields entering the standard model (left-handed leptons, i.e.\ neutrinos, in our case) will yield an additional term, the NY term. The both induced terms, i.e.\ the EH term and the NY one, are UV cutoff dependent, as usually in such cases, but fortunately, the BI parameter is not. It depends only on the number and kind of particle species entering the standard model. From purely technical point of view the calculus is partially akin to the derivation of the Adler--Bell--Jackiw (ABJ) chiral anomaly in space-time with torsion \cite{Zanelli}. More precisely the NY term dominates the anomaly (formally, it yields a divergence, which complicates a bit the calculation of the ABJ anomaly).

According to our realization of the Sakharov idea, we are interested in a dominant part of one-loop contributions coming from left-handed leptons. We will work in the (euclidean) Schwinger proper-time formalism and in the framework of the Seeley--DeWitt heat-kernel expansion on manifolds with torsion \cite{Obukhov}. Our starting object is the Dirac differential operator
\begin{align}
D\equiv i\! \not\!\nabla\equiv i\gamma^{a}e^{\mu}_{a}\,\nabla_{\mu},
\label{eq: Dirac differential operator}
\end{align}
where $e^{\mu}_{a}$ is a vierbein field, $\nabla_{\mu}$ is a covariant derivative in space with torsion, and $\gamma^{a}$ are euclidean Dirac matrices. Now
\begin{align}
D^{2}=-\square+\frac{1}{2}e^{\mu}_{a}e^{\nu}_{b}\sigma^{ab}T^{\lambda}_{\mu\nu}\nabla_{\lambda}-\frac{1}{8}e^{\mu}_{a}e^{\nu}_{b}\sigma^{ab}\sigma^{cd}R_{cd\mu\nu},
\label{eq: Dirac differential operator squared}
\end{align}
where
\begin{align}
\square\equiv\nabla_{\mu}\nabla^{\mu},\;\;\sigma^{ab}\equiv\frac{1}{2}\left[\gamma^{a},\gamma^{b}\right],\;\;\left[\nabla_{\mu}\, ,\nabla_{\nu}\right]V^{a}=R^{a}_{\;b\mu\nu}V^{b}-
T^{\lambda}_{\mu\nu}\nabla_{\lambda}V^{a}.
\label{eq: Dirac differential operator squared-notation}
\end{align}
Introducing the two chiral projectors
\begin{align}
P_{\rm \textsc{l}}\equiv\frac{1-\gamma^{5}}{2},\;\;P_{\rm \textsc{r}}\equiv\frac{1+\gamma^{5}}{2},
\label{eq: Projectors with gamma5}
\end{align}
with $\gamma^{5}\equiv\gamma^{1}\gamma^{2}\gamma^{3}\gamma^{4}$, we can write in the chiral representation
\begin{align}
D^{2}=D^{2}P_{\rm \textsc{l}}\oplus D^{2}P_{\rm \textsc{r}},
\label{eq: Dirac differential operator squared-simple sum}
\end{align}
and consequently
\begin{align}
\det D=\sqrt{\det D^{2}}=\sqrt{{\det}_{\rm \textsc{l}} D^{2}\,{\det} _{\rm \textsc{r}} D^{2}},
\label{eq: Dirac differential operator determinant}
\end{align}
because $D^{2}$ is diagonal-blocked in the subspaces L and R.
From now on we will confine ourselves to $\sqrt{{\det}_{\rm \textsc{l}} D^{2}}$ corresponding to the left-handed lepton.\\
The effective action for the chiral (left-handed) lepton is of the following form
\begin{align}
S=-\frac{1}{2}\log{\det}_{\rm \textsc{l}} D^{2}=\frac{1}{2}\int\frac{\mathrm{d}s}{s}\mathrm{Tr}\left(e^{-sD^{2}}P_{\rm \textsc{l}}\right).
\label{eq: Effective action for the chiral lepton}
\end{align}
Then, the chiral part of the $M^2$-regularized effective lagrangian density reads \cite{Zanelli} (there is a misprint in the coefficient in front of the NY term in the first reference of \cite{Zanelli}---that coefficient is twice bigger than in the second reference)
\begin{align}
\mathcal{L}=\frac{1}{2}\int\limits_{M^{-2}}^{\infty}\frac{\mathrm{d}s}{s}\,\frac{s}{(4\pi s)^{2}}\,\left(-\frac{1}{2}\right)\mathrm{tr}\left(a_{1}\gamma^{5}\right)=-\frac{1}{4}\left(\frac{M}{4 \pi}\right)^{2}\mathcal{NY}+O\left(\log M\right),
\label{eq: Effective Lagrangian for the chiral lepton}
\end{align}
where $a_{1}$ is the 1st Seeley--DeWitt coefficient \cite{Obukhov}, and the NY term $\mathcal{NY}$ is defined by
\begin{align}
\mathcal{NY}\equiv\mathrm{d}_{\omega}e^{a}\wedge\mathrm{d}_{\omega}e_{a}-e^{a}\wedge e^{b}\wedge R_{ab}\equiv T^{a}\wedge T_{a}-e^{a}\wedge e^{b}\wedge R_{ab}.
\label{eq: Nieh-Yan term definition}
\end{align}
It appears, as mentioned earlier, that the NY term $\mathcal{NY}$ dominates the chiral anomaly in space with torsion, i.e.\ \cite{Zanelli}
\begin{align}
\partial_{\mu}\!\left<J^{\mu}_{5}\right>=\left(\frac{M}{2\pi}\right)^{2}\mathcal{NY}+O(1),
\label{eq: Nieh-Yan domination}
\end{align}
where $O(1)$ means terms of the zeroth order in $M$. Due to that coincidence we could utilize Eq.~\eqref{eq: Nieh-Yan domination} for our purposes (i.e.\ in \eqref{eq: Effective Lagrangian for the chiral lepton}).\\
The extended lagrangian density of general relativity assumes the form
\begin{align}
\mathcal{L}=\alpha\star\left(e^{a}\wedge e^{b}\right)\wedge R_{ab}-\beta\left(T^{a}\wedge T_{b}-e^{a}\wedge e^{b}\wedge R_{ab}\right),
\label{eq: Extended lagrangian}
\end{align}
where the first term is the standard EH one, and the second term is the extended Holst or the NY one. The Barbero--Immirzi parameter $\gamma$ is now given by
\begin{align}
\gamma\equiv\frac{\alpha}{\beta}.
\label{eq: Barbero-Immirzi parameter}
\end{align}
Using the result of \cite{Broda1} we have
\begin{align}
\mathcal{L}_{\rm \textsc{eh}}=-\frac{1}{12}\,\left(\frac{M}{4\pi}\right)^{2}\left(N_{0}+N_{\frac{1}{2}}-4N_{1}\right)\star\left(e^{a}\wedge e^{b}\right)\wedge R_{ab},
\label{eq: E-H Lagrangian}
\end{align}
where $N_{0}$ is the number of minimal scalar degrees of freedom (dof), $N_{\frac{1}{2}}$ is the number of two-component fermion fields, and $N_{1}$ is the number of gauge fields (half the number of gauge dof).
Therefore, by virtue of \eqref{eq: Effective Lagrangian for the chiral lepton}, \eqref{eq: Nieh-Yan term definition} and \eqref{eq: Extended lagrangian}--\eqref{eq: E-H Lagrangian}
\begin{align}
\gamma=\frac{-\frac{1}{12}\left(N_{0}+N_{\frac{1}{2}}-4N_{1}\right)}{-\frac{1}{4}N_{\rm \textsc{l}}},
\label{eq: Induced gamma parameter}
\end{align}
where $N_{\rm \textsc{l}}$ is the number of chiral left-handed modes, and the UV cutoffs $\left(M/4 \pi \right)^{2}$ canceled out in \eqref{eq: Induced gamma parameter}.\\
For example, exactly in the framework of the standard model, we insert the following numbers of fundamental modes: $N_{0}=4$ (Higgs), $N_{\frac{1}{2}}=45$, $N_{1}=12$, $N_{\rm \textsc{l}}=3$ (neutrinos), yielding $\gamma=\frac{1}{9}\approx 0.11$, which is quite close to the (a bit obsolete) Ashtekar--Baez--Corichi--Krasnov value, $\gamma_{\!\rm\textsc{abck}}=\frac{\ln2}{\pi\sqrt{3}}\approx 0.13$ \cite{Baez},\cite{Domagala} (see \cite{Meissner}, for a better estimation). Nevertheless we should remember that $\gamma$ induced that way depends on the number and kinds of fundamental modes, and moreover the whole calculus is valid in the framework of euclidean formalism. One should also note that right-handed fermions would yield $\gamma=-\frac{1}{9}$.\\
One could ask a question what is the status of the result obtained in this letter. First of all, one should observe that the both parts of the gravitational action, i.e.\ the standard EH part and the Holst (or the NY) part, can be derived in a uniform way. Namely, we can (quantumly) ``induce'' them, including coefficient, from matter fields entering the standard model. Therefore, we should treat the derivation of the Holst (or the NY) term the same way we treat the induced gravity. In Visser's terminology \cite{Visser}, we assume the Sakharov's one-loop dominance interpretation \cite{Sakharov} (all tree-level constants set to zero). Obviously, it is only a technical side of our derivation. As far as a conceptual side is concerned, we would like to cite our previous work \cite{Broda1}: ``Actually, at present, the very idea lacks a clear theoretical interpretation. It can be treated either as an interesting curiosity or as an unexplained deeper phenomenon. Anyway, coincidences are striking. Our point of view is purely pragmatical ...''. Moreover at the moment, we do not see any conceptual nor technical relationship between our approach and the ``standard'' approach using loop quantum gravity and black-hole entropy \cite{Domagala},\cite{Baez},\cite{Meissner}.

\section*{Acknowledgments}
We would like to thank Simone Mercuri for his e-mail.\\
This work has been supported by the University of {\L}\'od\'z
grant and LFPPI network.

\end{document}